\documentclass[aps,twocolumn,superscriptaddress]{revtex4-2}

\usepackage[T1]{fontenc}

\usepackage{amsmath,amssymb,amsfonts,color,graphicx,tabularx}

\usepackage[unicode=true,colorlinks=true]{hyperref}

\hypersetup{linkcolor=blue,citecolor=blue,urlcolor=blue}

\newcommand{\overbar}[1]{\overline{\mkern-2.2mu#1\mkern-0.2mu}}

\begin{document}

\title{Numerical extraction of crosscap coefficients in microscopic models \\ for (2+1)D conformal field theory}

\author{Jia-Ming Dong}
\thanks{These two authors contributed equally.}
\affiliation{School of Physics and Wuhan National High Magnetic Field Center, Huazhong University of Science and Technology, Wuhan 430074, China}

\author{Yueshui Zhang}
\thanks{These two authors contributed equally.}
\affiliation{Faculty of Physics and Arnold Sommerfeld Center for Theoretical Physics, Ludwig-Maximilians-Universit\"at M\"unchen, 80333 Munich, Germany}

\author{Kai-Wen Huang}
\affiliation{School of Physics and Wuhan National High Magnetic Field Center, Huazhong University of Science and Technology, Wuhan 430074, China}

\author{Hong-Hao Tu}
\email{h.tu@lmu.de}
\affiliation{Faculty of Physics  and Arnold Sommerfeld Center for Theoretical Physics, Ludwig-Maximilians-Universit\"at M\"unchen, 80333 Munich, Germany}

\author{Ying-Hai Wu}
\email{yinghaiwu88@hust.edu.cn}
\affiliation{School of Physics and Wuhan National High Magnetic Field Center, Huazhong University of Science and Technology, Wuhan 430074, China}

\begin{abstract}
Conformal field theory (CFT) can be placed on disparate space-time manifolds to facilitate investigations of their properties. For (2+1)-dimensional [(2+1)D] theories, one useful choice is the real projective space $\mathbb{RP}^3$ obtained by identifying antipodal points on the boundary sphere of a three-dimensional ball. One-point functions of scalar primary fields on this manifold generally do not vanish and encode the so-called crosscap coefficients. These coefficients also manifest on the sphere as the overlaps between certain crosscap states and CFT primary states. Taking the (2+1)D Ising CFT as a concrete example, we demonstrate that crosscap coefficients can be extracted from microscopic models. We construct crosscap states in both lattice models defined on polyhedrons and continuum models in Landau levels, where the degrees of freedom at antipodal points are entangled in Bell-type states. By computing their overlaps with the eigenstates of many-body Hamiltonians, we obtain results consistent with those from conformal bootstrap. Importantly, our approach directly reveals the absolute values of crosscap overlaps, whereas bootstrap calculations typically yield only their ratios. Furthermore, we investigate the finite-size scaling of these overlaps and their evolution under perturbations away from criticality.
\end{abstract}

\maketitle

{\it Introduction} --- 
Phase transitions are intriguing phenomena that arise from the interplay of a large number of constituents in many-body systems. At the critical points in many continuous transitions, conformal symmetry may emerge even if it does not exist at the microscopic level~\cite{Polyakov1970}. Building upon this feature, the powerful framework of conformal field theory (CFT) was established~\cite{Belavin1984}. Besides critical phenomena, CFTs have also been applied to study particle physics and quantum gravity. From a mathematical perspective, CFTs can be defined rigorously unlike general quantum field theory. In two-dimensional classical systems or (1+1)-dimensional [(1+1)D] quantum systems, local conformal transformations are described by an infinite-dimensional algebra so they place tight constraints on physical properties. This allows for non-perturbative analytical calculations of many interesting quantities.

As we turn to higher spatial dimensions, CFTs become much more difficult because the associated algebra is finite dimensional~\cite{Rychkov-Book}. One successful approach is conformal bootstrap in which correlation functions are analyzed by imposing suitable constraints derived from conformal symmetry~\cite{ElShowk2012,Poland2019}. Another approach is the direct numerical simulations of microscopic models, which have also played a crucial role in studying CFTs. This method faces challenges about how to design suitable models, how to determine if there are continuous transitions, and how to extract the CFT data at critical points~\cite{DengYJ2002,Schuler2016}. In the past three years, an important progress has been made in quantum critical model design and extraction of CFT data~\cite{ZhuW2023}. Its essence is to put electrons on a sphere with a magnetic monopole at its center~\cite{WuTT1976,WuTT1977,Haldane1983c}. The single-particle eigenstates on the sphere are organized into Landau levels (LLs) each having a finite degeneracy. If the cyclotron gap is sufficiently large, the electrons can be restricted to the lowest LL, which make numerical calculations feasible. More importantly, spherical symmetry is preserved exactly so energy spectra can be analyzed using the state-operator correspondence~\cite{Cardy1984a,Cardy1985a}. This idea has been employed and further developed in subsequent works~\cite{HuLD2023,HanC2023,HuLD2024,HanC2024,Hofmann2024,ZhouZ2024b,HuLD2025,Fardelli2025,FanRH2024,ZhouZ2025,Dedushenko2024,ZhouZ2024a,ChenBB2024,Voinea2025,YangS2025a,Lauchli2025,FanRH2025,Cruz2025,Miro2025,HeYC2025,Taylor2025,YangS2025b}. For the three-dimensional (3D) Ising CFT, numerical results obtained by this approach and conformal bootstrap are remarkably close to each other (when both are available). Another related progress is the identification of the Gross-Neveu-Yukawa criticality in interacting Dirac fermions on the sphere without magnetic monopole~\cite{GaoZQ2025}.  
 
\begin{figure}[ht]
\includegraphics[width=0.48\textwidth]{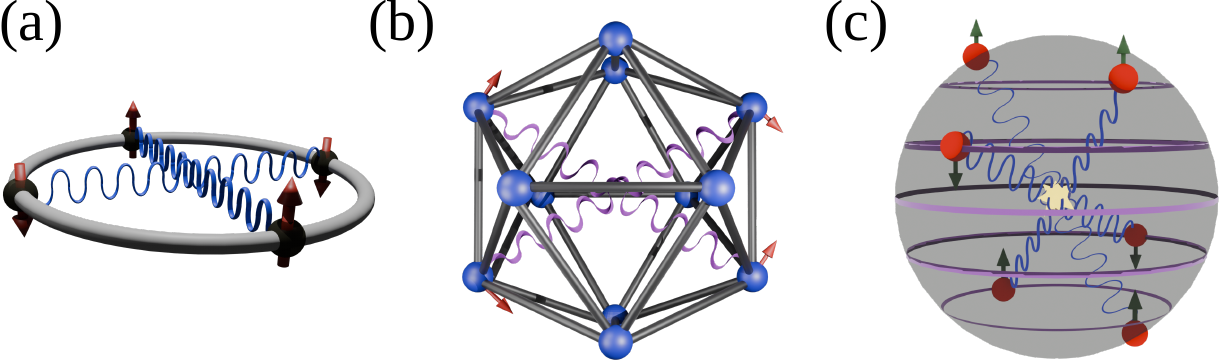}
\caption{Schematics of crosscap states in (a) one-dimensional systems, (b) the icosahedron, (c) LLs on the sphere.}
\label{Figure1}
\end{figure}

Shortly after the inception of CFT, it was recognized that global topological features, such as boundary conditions or nontrivial manifolds, can lead to rich and universal phenomena in finite-size systems~\cite{Cardy1984b,Cardy1986,Cardy1989,IshibashiN1989,Affleck1991a}. One way to explore their effects is through the study of conformal boundary states. In this work, we focus on one special class of conformal boundary states, known as crosscap states~\cite{Blumenhagen-Book,Fioravanti1994,Pradisi1996,Fuchs2000,Brunner2004,ChoGY2015,Caetano2022,Ekman2022,Gombor2022,Gombor2023,HeM2023,ChibaY2024,YonetaY2024,TanBY2025,ZhangYS2024,Chalas2024,WeiZX2025,WeiZX2024,LiYZ2025,Mohapatra2025}. The name stems from their analogy with the geometric object crosscap and is most easily visualized in one spatial dimension. A crosscap is obtained from a circle by gluing together both ends of each diameter. For one-dimensional (1D) lattice or continuum models, crosscap states are constructed by putting each pair of antipodal points into entangled states [see Fig.~\ref{Figure1} (a)]. This definition extends to integrable models that are not described by CFTs. In some cases, they are called entangled antipodal states~\cite{ChibaY2024,YonetaY2024}. An intriguing feature of the crosscap states is that their overlaps with some CFT eigenstates are universal. In this paper, we construct crosscap states for (2+1)-dimensional [(2+1)D] CFT and perform numerical calculations to obtain their overlaps in finite-size systems. We mainly focus on the spherical geometry but shall also consider polyhedrons~\cite{LaoBX2023}. 

{\it Crosscap overlap in (2+1)D CFT} --- We consider a (2+1)D CFT with the quantum Hamiltonian defined on the surface of a sphere with radius $R$. The space-time manifold is $\mathbb{R} \times S^2 $, with coordinate $(v\tau,\vec{x})$, where $v$ is the velocity and $\vec{x}$ denotes the spatial coordinate constrained to the spherical surface $|\vec{x}|=R$. The spherical surface can be parametrized as $\vec{x} = R \,\vec{\omega}$, where $\vec{\omega}=(\sin\theta \cos\phi, \sin\theta \sin\phi, \cos\theta)$ is a unit vector specified by the angular coordinates $\theta$ and $\phi$. The radial quantization scheme is realized via the conformal transformation $\vec{r} = e^{v\tau/R}\,\vec{\omega}$~\cite{Cardy1985a}, which maps the space-time manifold to the flat Euclidean space $\mathbb{R}^{3}$. Under this transformation, the spatial manifold at the $\tau=0$ time slice becomes a unit sphere. In this scheme, there is a state-operator correspondence: each CFT primary state $|\mathcal{O}_{i}\rangle$ is associated with a scalar primary field $\mathcal{O}_{i}(\vec{r})$ with $\vec{r}$ being the space-time coordinate~\cite{Cardy1984a}. We aim to construct crosscap states $|\mathcal{C}\rangle$ on the sphere and compute the crosscap overlaps between them and the CFT eigenstates. The physical picture for 1D crosscap states shall provide valuable guidance for this quest. For the primary states, state-operator correspondence tells us that
\begin{align}
\langle \mathcal{C} | \mathcal{O}_{i}\rangle = \lim_{|\vec{r}|\to 0} \langle \mathcal{C} | \mathcal{O}_{i}(\vec{r}) | 0 \rangle \,, 
\label{eq:cross-ovlp}
\end{align}
where the CFT vacuum is denoted as $|0\rangle$. In the path integral formalism, the crosscap state is associated with a boundary condition that identifies antipodal points on the surface of a unit ball. This identification converts the space-time manifold to the 3D real projective space $\mathbb{RP}^{3}$. Many properties about CFTs on $\mathbb{RP}^{3}$ have been studied before~\cite{NakayamaY2016,Andrei2020,Giombi2021}. The crosscap overlap in Eq.~\eqref{eq:cross-ovlp} can be reinterpreted as the one-point function
\begin{align}
\langle \mathcal{C} | \mathcal{O}_{i} \rangle = \langle \mathcal{O}_{i}(\vec{r}=\vec{0}) \rangle_{\mathbb{RP}^{3}}
\end{align}
on the $\mathbb{RP}^{3}$ manifold, with primary field $\mathcal{O}_{i}(\vec{r})$ inserted at the origin.

Since translation symmetry is explicitly broken, the one-point functions of primary fields generally do not vanish on $\mathbb{RP}^{3}$. In fact, their forms are fully constrained by conformal transformations generated by $\vec{P}+\vec{K}$ as they preserve the geometry of $\mathbb{RP}^3$. Here $\vec{P}$ and $\vec{K}$ respectively denote the generators of translations and special conformal transformations, whose explicit expressions are given by
\begin{align}
    \vec{P} &= -i\frac{\partial}{\partial \vec{r}}\, , \nonumber\\
    \vec{K} &= -i \left[2\vec{r}\, \left(\vec{r}\cdot \frac{\partial}{\partial \vec{r}}\right) - |\vec{r}|^{2} \frac{\partial}{\partial \vec{r}}\right]\, .
\end{align}
It is helpful to study the following finite conformal transformation:
\begin{align}
\vec{f}(\vec{r}) &= e^{i\vec{a}\cdot (\vec{P}+\vec{K})}(\vec{r}) \nonumber\\
                 &= \frac{1 + (\tan |\vec{a}|)^{2} }{1+ \left(|\vec{r}|\tan |\vec{a}|\right)^{2} -2 \tan |\vec{a}|\, \frac{(\vec{r}\cdot\vec{a})}{|\vec{a}|}}\, \vec{r}\nonumber\\
                 & \quad +\frac{(1-|\vec{r}|^{2}) \tan |\vec{a}| -2\tan^{2} |\vec{a}|\, \frac{(\vec{r}\cdot\vec{a})}{|\vec{a}|}}{1+ \left(|\vec{r}|\tan |\vec{a}|\right)^{2} -2 \tan |\vec{a}|\, \frac{(\vec{r}\cdot\vec{a})}{|\vec{a}|}}\, \frac{\vec{a}}{|\vec{a}|}\, .
\end{align}
Under this conformal transformation, the origin is mapped to $\vec{r}\equiv \vec{f}(\vec{0}) = \tan |\vec{a}|\, \frac{\vec{a}}{|\vec{a}|}$, and a scalar primary field transforms as $\mathcal{O}_{i} (\vec{r}) = |\vec{f}'(\vec{0})|^{-\Delta_{i}/3} \mathcal{O}_{i} (\vec{0})$, where $\Delta_{i}$ is the scaling dimension of this field and $|\vec{f}'(\vec{0})| = (1 + \tan^{2} |\vec{a}|)^{3}$ is the Jacobian determinant of the conformal transformation. One thus concludes that one-point functions on $\mathbb{RP}^3$ must take the form
\begin{align}
\langle\mathcal{O}_{i}(\vec{r})\rangle_{\mathbb{RP}^{3}} = \frac{\mathcal{A}_{i}}{(1+|\vec{r}|^{2})^{\Delta_{i}}}
\end{align}
with $\mathcal{A}_{i}$ being a universal coefficient that is identified with the crosscap overlap
\begin{align}
\mathcal{A}_{i} = \langle \mathcal{C}| \mathcal{O}_{i} \rangle\, .
\label{eq:cross-coeff}
\end{align}
Here, $\mathcal{A}_{i}$ is the so-called crosscap coefficient.

For (1+1)D rational CFTs, the crosscap coefficients encode conformal data on flat space and can be computed analytically~\cite{Fioravanti1994,Pradisi1995,Fuchs2000}. Their values have also been extracted via crosscap overlaps in a variety of microscopic models~\cite{TuHH2017,TangW2017,ChenL2017,TangW2019,TangW2020,LiZQ2020,VanHove2022,ZhangYS2023,Shimizu2024,TanBY2025,ZhangYS2024,LiuT2025}. In contrast, analytical results about crosscap coefficients in (2+1)D CFTs are generally unavailable. Nevertheless, the \emph{ratios} between crosscap coefficients of some primary fields and that of the identity primary field have been obtained by numerical calculations based on the conformal bootstrap approach~\cite{NakayamaY2016}. Instead, absolute values of the crosscap coefficients -- not just their ratios -- can be extracted directly from the crosscap overlaps in Eq.~\eqref{eq:cross-coeff}. The feasibility of this approach is demonstrated below using concrete microscopic models that realize the (2+1)D Ising CFT. For quantum many-body Hamiltonians that are defined on the sphere or its discrete counterpart, low-lying eigenstates can be obtained by exact diagonalization. It is not {\em a priori} clear how to construct crosscap states in these models. According to the intuitive picture mentioned above, the degrees of freedom on the two ends of each diameter of the sphere should form a maximally entangled state. 

{\it Lattice models on icosahedron} --- We begin with a two-dimensional quantum Ising model defined on discrete lattice sites. The intuition accumulated during the past about 1D crosscap states can be extended easily to these cases. However, the usual planar lattice models with or without periodic boundary conditions are not suitable for studying crosscap states. Instead, highly symmetric polyhedrons may be considered~\cite{LaoBX2023}. As shown in Fig.~\ref{Figure1} (b), the transverse-field Ising (TFI) model
\begin{eqnarray}
H_{\rm TFI} = - \sum_{\langle{ij}\rangle} \sigma^{x}_{i} \sigma^{x}_{j} - h \sum_{i} \sigma^{z}_{i}
\end{eqnarray}
is defined on the 12 vertices of icosahedron. An inspection of the excited state energy levels suggests that $h=4.375$ is the critical point and several primary states can be pinpointed~\cite{LaoBX2023}. We are interested in crosscap overlaps of the ground state, the $\epsilon$ primary state, and the $\epsilon'$ primary state, which are denoted as $\mathcal{A}_{\mathbb{I}}$, $\mathcal{A}_{\epsilon}$, and $\mathcal{A}_{\epsilon'}$, respectively. It is possible to define a point as the center of icosahedron that has the same distance to all vertices. This allows one to construct a fictitious sphere such that the vertices can be divided into six groups each having two vertices on a diameter of the fictitious sphere. For each group with two vertices denoted as $i$ and $\overbar{i}$, the Bell state $|\uparrow\rangle_{i}|\uparrow\rangle_{\overbar{i}} + |\downarrow\rangle_{i}|\downarrow\rangle_{\overbar{i}}$ is imposed. It is natural to accept that
\begin{eqnarray}
|\mathcal{C}_{\rm IC}\rangle = \prod_{i\in{\rm half}} \left( |\uparrow\rangle_{i}|\uparrow\rangle_{\overbar{i}} + |\downarrow\rangle_{i}|\downarrow\rangle_{\overbar{i}} \right)
\end{eqnarray}
is a crosscap state for the whole system. This expression gives rise to $\mathcal{A}_{\mathbb{I}}=1.133$, $\mathcal{A}_{\epsilon}=0.782$, and $\mathcal{A}_{\epsilon'}=1.293$ at the critical point. Using the bootstrap method, Ref.~\cite{NakayamaY2016} found that $\mathcal{A}_{\epsilon}/\mathcal{A}_{\mathbb{I}}=0.667$ and $\mathcal{A}_{\epsilon'}/\mathcal{A}_{\mathbb{I}}=0.896$. The former value is not far from ours but the latter value is quite different. We summarize the results in Table~\ref{Table1} for comparison. 

\begin{table}[ht]
\begin{tabular}{c|ccc}
\hline
\hline
method   &  $\mathcal{A}_{\mathbb{I}}$ & $\mathcal{A}_{\epsilon}/\mathcal{A}_{\mathbb{I}}$ & $\mathcal{A}_{\epsilon'}/\mathcal{A}_{\mathbb{I}}$ \\
            \cline{2-4}          
\hline
bootstrap             &    NA   &  0.667  &  0.896  \\
icosahedron           &  1.133  &  0.690  &  1.141  \\
sphere (extrapolated) &  1.129  &  0.674  &  0.898  \\
\hline
\hline
\end{tabular}
\caption{Comparison between the crosscap coefficient results for the 3D Ising CFT in conformal bootstrap calculations and microscopic lattice and continuum models. NA means not available.}
\label{Table1}
\end{table}

\begin{figure}[ht]
\includegraphics[width=0.48\textwidth]{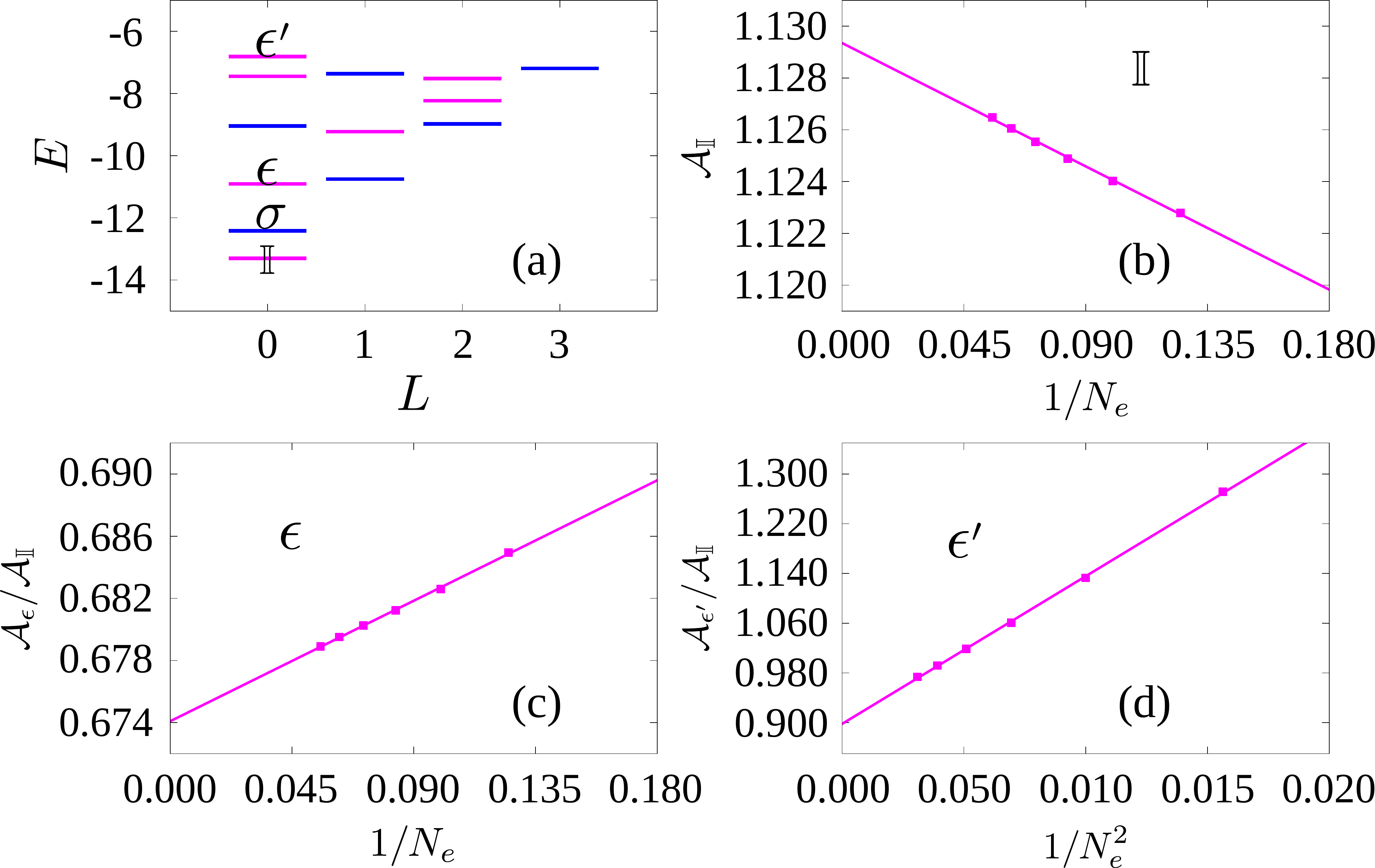}
\caption{Numerical results at the critical point. (a) Energy spectrum of the system with $N_{e}=18$. The ground state $\mathbb{I}$ and three primary states $\sigma,\epsilon,\epsilon'$ are indicated. Eigenstates with even and odd $Z_2$ parity are represented by magenta and blue colors, respectively. (b) Ground state crosscap overlaps. (c) Ratios between the $\epsilon$ state crosscap overlaps and the ground state crosscap overlaps. (d) Ratios between the $\epsilon'$ state crosscap overlaps and the ground state crosscap overlaps. The solid lines in panels (b-c) are linear fitting versus $1/N_{e}$ and the one in panel (d) is linear fitting versus $1/N^{2}_{e}$.}
\label{Figure2}
\end{figure}

{\it Continuum models in Landau levels} --- The results on icosahedron are quite impressive given the small number of sites. In this section, we turn to the main topic of this paper. We employ the model constructed in Ref.~\cite{ZhuW2023} to study Ising phase transitions on the sphere [see Fig.~\ref{Figure1} (c)]. The single-particle eigenstates are monopole harmonics~\cite{WuTT1976,WuTT1977}. If the magnetic fluxes through the sphere is $2Q$, explicit expressions for the lowest LL are
\begin{eqnarray}
Y_{Q,m}(\boldsymbol{\Omega}) &=& \left[ \frac{2Q+1}{4\pi} \binom{2Q}{Q-m} \right]^{\frac{1}{2}} (-1)^{Q-m} \exp\left(im\phi\right) \nonumber \\
&\phantom{=}& \times \cos^{Q+m}(\theta/2) \sin^{Q-m}(\theta/2),
\end{eqnarray}
where $\theta$ is the azimuthal angle, $\phi$ is the radial angle, $\boldsymbol{\Omega}=(\theta,\phi)$ is the solid angle, and $m$ is the $z$ component of angular momentum. There are $2Q+1$ Landau orbitals with $m \in [-Q,-Q+1,\cdots,Q-1,Q]$. In units of the magnetic length, the radius of the sphere is $R=\sqrt{Q}$. The electrons are endowed with an internal degree of freedom called spin with two possible values $\uparrow$ and $\downarrow$. For the Landau orbital with spin index $\sigma$ and angular momentum $m$, the annihilation (creation) operator is denoted as $C_{\sigma,m}$ ($C^{\dag}_{\sigma,m}$). In real space, their counterparts are $\Psi_{\sigma}(\boldsymbol{\Omega})=\sum_{m} Y_{Q,m}(\boldsymbol{\Omega}) C_{\sigma,m}$ and $\Psi^{\dag}_{\sigma}(\boldsymbol{\Omega})=\sum_{m} \left[ Y_{Q,m}(\boldsymbol{\Omega}) \right]^{*} C^{\dag}_{\sigma,m}$. The system has $N_{e}=2Q+1$ electrons that interact with each other via the two-body potential $V_{\sigma\tau}(\boldsymbol{\Omega}_{1},\boldsymbol{\Omega}_{2})$. It should satisfy the criterion: when this is the only term in the many-body Hamiltonian, the system has two degenerate ground states in which all electrons have the same spin index. This is the ferromagnetic side of the Ising transition. We choose $V_{\uparrow\uparrow}=V_{\downarrow\downarrow}=0$ and $V_{\uparrow\downarrow}(\boldsymbol{\Omega}_{1},\boldsymbol{\Omega}_{2})=g_{0} \delta(\boldsymbol{\Omega}_{1}-\boldsymbol{\Omega}_{2}) + g_{1} \nabla^{2} \delta(\boldsymbol{\Omega}_{1}-\boldsymbol{\Omega}_{2})$ as in Ref.~\cite{ZhuW2023}. The second-quantized form of this potential is~\footnote{The prefactor $4$ is chosen to match the convention in Ref.~\cite{ZhuW2023}.}
\begin{eqnarray}
H_{a} = 4 \sum_{\{m_{i}\}} V_{m_{1}m_{2}m_{3}m_{4}} C^{\dag}_{\uparrow,m_{1}} C^{\dag}_{\downarrow,m_{2}} C_{\downarrow,m_{4}} C_{\uparrow,m_{3}},
\end{eqnarray}
where the two-body matrix elements are expressed in terms of the Haldane pseudopotentials $V_{0},V_{1}$~\cite{Haldane1983c} and Clebsch-Gordon coefficients
\begin{widetext}
\begin{eqnarray}
V_{m_{1}m_{2}m_{3}m_{4}} &=& \int \mathrm{d}\Omega_{1} \int \mathrm{d}\Omega_{2} \left[ Y_{Q,m_{1}}(\boldsymbol{\Omega}_{1}) \right]^{*} \left[ Y_{Q,m_{2}}(\boldsymbol{\Omega}_{2}) \right]^{*} V_{\uparrow\downarrow}(\boldsymbol{\Omega}_{1},\boldsymbol{\Omega}_{2}) Y_{Q,m_{3}}(\boldsymbol{\Omega}_{2}) Y_{Q,m_{4}}(\boldsymbol{\Omega}_{1}) \nonumber \\
&=& \delta_{m_{1}+m_{2},m_{3}+m_{4}} V_{0} \langle Q,m_{1};Q,m_{2}| 2Q,m_{1}+m_{2}\rangle \langle 2Q,m_{3}+m_{4} | Q,m_{3};Q,m_{4}\rangle \nonumber \\
&\phantom{=}& + \; \delta_{m_{1}+m_{2},m_{3}+m_{4}} V_{1} \langle Q,m_{1};Q,m_{2}| 2Q-1,m_{1}+m_{2}\rangle \langle 2Q-1,m_{3}+m_{4} | Q,m_{3};Q,m_{4}\rangle.
\end{eqnarray}
\end{widetext}
To induce a phase transition, we introduce a tunneling term 
\begin{eqnarray}
H_{b} = -h \sum_{m} \left( C^{\dag}_{\uparrow,m} C_{\downarrow,m} + C^{\dag}_{\downarrow,m} C_{\uparrow,m} \right)
\end{eqnarray}
between spin-up and spin-down orbitals at the same angular momentum. The full Hamiltonian $H=H_{a}+H_{b}$ respects SO(3) rotational symmetry so its eigenstates can be labeled by total angular momentum $L(L+1)$. In the limit of $h{\rightarrow}\infty$, each electron is driven into an equal superposition of spin-up and spin-down. This is the paramagnetic side of the Ising transition. The critical point with minimal finite-size effects was located at $V_{0}=4.75$, $V_{1}=1$, and $h_{c}=3.16$~\cite{ZhuW2023}. Exact diagonalization has been performed in multiple systems with up to $N_{e}=18$ electrons. The energy spectrum of the largest system at the critical point is displayed in Fig.~\ref{Figure2} (a). We have identified three primary states $\sigma$, $\epsilon$, and $\epsilon'$ following the procedure described in Ref.~\cite{ZhuW2023}.

It is not immediately clear how to define crosscap states on the sphere at the microscopic level. We conjecture that
\begin{eqnarray}
|\mathcal{C}_{\rm LL}\rangle &=& \exp \Big\{ \int_{\rm upper} \mathrm{d}\boldsymbol{\Omega} \; \left[ \Psi^{\dag}_{\uparrow}(\theta,\phi) \Psi^{\dag}_{\uparrow}(\pi-\theta,\pi+\phi) \right. \nonumber \\
&& + \left. \Psi^{\dag}_{\downarrow}(\theta,\phi) \Psi^{\dag}_{\downarrow}(\pi-\theta,\pi+\phi) \right] \Big\} |\emptyset\rangle
\label{eq:cross_sphere}
\end{eqnarray}
is a crosscap state, where the integral is restricted to the upper half-sphere and $|\emptyset\rangle$ is the vacuum state. While $\phi+\pi$ could be larger than $2\pi$, periodicity of the monopole harmonics ensures that they are still well defined. To mimic 1D circles are split as two semicircles, we divide the sphere into the upper and lower halves. It is possible to consider other bipartitions but the present choice is the most convenient one. For the point with coordinates $(\theta,\phi)$ in the upper half, a diameter of the sphere connects it with another point $(\pi-\theta,\pi+\phi)$ on the lower half. Inside the square bracket, the combination of operators indicates that electrons are created at these two points with their spins forming a Bell state. We cannot multiply these entangled states together due to the continuous nature of the sphere. Instead, putting these operators in the exponent leads to a well-defined state. For a crosscap state in continuum field theory, a certain part of the conformal symmetry should be preserved. However, the state proposed above does not have this property. This is not surprising because the spherical microscopic model does not have exact conformal symmetry. It only emerges in the thermodynamic limit at the critical point. If a Taylor expansion is performed on Eq.~\eqref{eq:cross_sphere}, the $M$th-order term would contain $2M$ electrons that are decomposed to $M$ pairs of maximally entangled states. It could have nonzero overlaps with many-body eigenstates when $2M=N_{e}$.

When the real-space operators are projected to the lowest Landau level, the proposed state becomes
\begin{eqnarray}
|\mathcal{C}_{\rm LL}\rangle = \exp \left\{ \sum_{\sigma} \sum_{m,n} F^{*}_{m,n} C^{\dag}_{\sigma,m} C^{\dag}_{\sigma,n} \right\} |\emptyset\rangle.
\end{eqnarray}
The coefficient $F_{m,n}$ can be evaluated as
\begin{eqnarray}
&& \int^{\pi/2}_{0} \sin\theta \; \mathrm{d}\theta \int^{2\pi}_{0} \mathrm{d}\phi \; Y_{Q,m}(\theta,\phi) Y_{Q,n}(\pi-\theta,\pi+\phi) \nonumber \\
= && \; \delta_{m,-n} (-1)^{2Q} e^{-im\pi} I_{1/2}(Q-m+1,Q+m+1)
\end{eqnarray}
with $I_{x}(a,b)$ being the regularized incomplete beta function. If $2Q+1$ is even, the crosscap state is
\begin{eqnarray}
|\mathcal{C}_{\rm LL}\rangle = \exp \left[ \sum_{\sigma} \sum_{m>0} (-1)^{m+1} C^{\dag}_{\sigma,m} C^{\dag}_{\sigma,-m} \right] |\emptyset\rangle \, .
\end{eqnarray}
Its fermion parity is even so there could be nonzero overlaps with the ground state, $\epsilon$ state, and $\epsilon'$ state. This is indeed the case as confirmed by explicit calculations. We show in Fig.~\ref{Figure2} (b) the overlaps for the ground states. For the 1D TFI model at its critical point, there is no finite-size correction to the crosscap overlaps~\cite{ZhangYS2024}. In contrast, the crosscap overlaps on the sphere varies with $N_{e}$, but they can be fitted very well using a linear function of $1/N_{e}$. It yields $1.129$ in the thermodynamic limit, which is quite close to the value on icosahedron. For the $\epsilon$ and $\epsilon'$ states, crosscap overlaps for each $N_{e}$ divided by the ground-state overlap in the same system are displayed in Fig.~\ref{Figure2} (c-d). These numbers can be compared directly with the bootstrap results~\cite{NakayamaY2016}. Finite-size effects for the $\epsilon$ state is quite weak, and the result is already quite close to $0.667$ even at the smallest system. When a linear fitting versus $1/N_{e}$ is performed on the data, we obtain $0.674$ in the thermodynamic limit. The ratios for the $\epsilon'$ state exhibit strong variations as $N_{e}$ increases. For the largest system with $N_{e}=18$, its value $0.974$ is still quite different from the bootstrap result. Linear fitting of finite-size data versus $1/N_{e}$ does not work well. Instead, the coefficient of determination is $0.9996$ for linear fitting versus $1/N^{2}_{e}$ and the extrapolated number is $0.898$. We stress that linear or quadratic fittings for the crosscap overlaps and ratios are purely empirical without theoretical underpinning at present.

\begin{figure}[ht]
\includegraphics[width=0.48\textwidth]{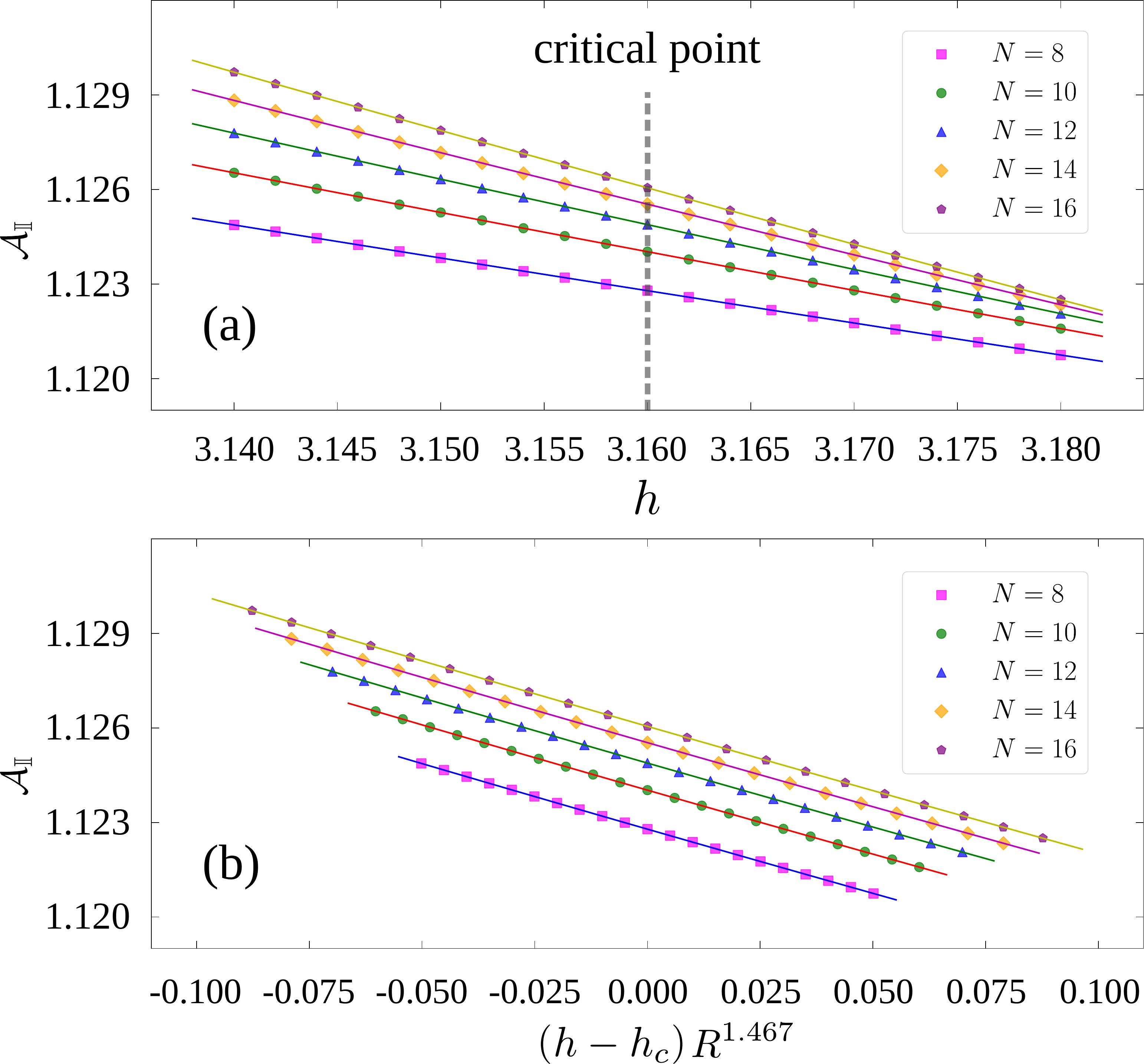}
\caption{Numerical results off the critical point. (a) Ground state crosscap overlaps for $h\in[3.14,3.18]$. (b) The same data in panel (a) plotted versus $(h-h_{c})R^{1.467}$. The solid lines in both panels are linear fitting results. } 
\label{Figure3}
\end{figure}

For a (1+1)D CFT perturbed by a relevant operator, crosscap overlap of the ground state was shown to be a universal scaling function of the dimensionless coupling related to this operator~\cite{ZhangYS2023}. This observation can be extended to (2+1)D systems in a straightforward manner. Let us consider the Hamiltonian
\begin{align}
    H = H_{\mathrm{CFT}} - v g \int_{|\vec{x}| = R} \mathrm{d}\vec{x} \, \varphi(\vec{x})
\label{eq:perturbed-H}
\end{align}
defined on a sphere with radius $R$. $H_{\mathrm{CFT}}$ is the unperturbed CFT Hamiltonian, $\varphi$ is an operator with scaling dimension $\Delta_{\varphi}<3$ for it to be relevant in renormalization group flow, $v$ is the velocity (encoded in $H_{\mathrm{CFT}}$), and $g$ is the coupling constant. It is apparent that $H_{\mathrm{CFT}}$ has dimension $[v/R]$ and $\varphi$ has dimension $[R]^{-\Delta_{\varphi}}$, so $g$ must have dimension $[R]^{\Delta_{\varphi}-3}$ to match the two terms in Eq.~\eqref{eq:perturbed-H}. We thus conclude that the whole theory is governed by a single \emph{dimensionless} coupling $s = g R^{3-\Delta_{\varphi}}$ and its ground state can be written as $|G(s)\rangle$. The crosscap overlap $\langle \mathcal{C}|G(s)\rangle$ only depends on the dimensionless coupling and can be captured by a universal scaling function. In the spherical Ising model, the tunneling term becomes a relevant perturbation when $h$ is tuned away from the critical point. For many cases with $h\in[3.14,3.16]$, the ground-state crosscap overlaps have been evaluated. As one can see from Fig.~\ref{Figure3} (a), the values in each system are remarkably linear versus $h-h_{c}$. Based on these results, we conjecture that crosscap coefficients change monotonically when relevant perturbations are turned on to induce a spontaneous symmetry breaking phase transition. For each fitting line in Fig.~\ref{Figure3} (a), its slope is denoted as $\mu(N_{e}) \equiv \mu(R)$. These values can be fitted very well using the formula $\mu(R)=0.041R^{1.467}$. It is instructive to replot the overlaps versus $(h-h_c)R^{1.467}$ as the fitting lines would become parallel to each other [see Fig.~\ref{Figure3} (b)]. From the CFT perspective, the system in the infrared limit is described by the Ising CFT perturbed by the $\epsilon$ primary field (``thermal perturbation''). According to our general analysis, the overlaps should depend on a dimensionless coupling that is proportional to $(h-h_c)R^{3-\Delta_{\epsilon}}$. This exponent is $1.587$ since bootstrap calculations found $\Delta_\epsilon \approx 1.413$~\cite{ElShowk2012}, which is reasonably close to our numerical result.

{\it Summary and outlook} --- To sum up, crosscap states have been constructed in microscopic models for the (2+1)D Ising CFT and extracted crosscap coefficients using their overlaps. The results are consistent with the ratios between crosscap coefficients obtained by conformal bootstrap calculations~\cite{NakayamaY2016}. We hope that the validity of our crosscap state can be further substantiated by additional investigations. Since the numerical values depend on the system size, it would be useful if larger numbers of electrons can be accessed by other methods such as density matrix renormalization group~\cite{White1992}. On the analytical side, direct calculations about the finite-size behaviors of crosscap coefficients are obviously desirable. In a previous work by some of the authors, conformal perturbation theory was developed for (1+1)D CFTs to understand finite-size corrections to crosscap overlaps~\cite{ZhangYS2023}. It was found that irrelevant and marginal terms generally lead to different types of finite-size corrections. If similar results can be established in (2+1)D CFTs, data fitting would be much more reliable. Besides the crosscap coefficients, one- and two-point correlation functions associated with the proposed crosscap state should also be studied. The results can be compared with previous results obtained by conformal bootstrap~\cite{NakayamaY2016}.

Many other problems about crosscap states and coefficients are largely unclear and deserve further investigations. It would be interesting to explore if bootstrap or other methods can also yield the absolute values of the crosscap overlaps. For several other paradigmatic (2+1)D CFTs, little is known about the crosscap coefficients. In particular, do they encode certain physical quantities? For the (1+1)D compactified boson CFT, which is the low-energy theory of Tomonaga-Luttinger liquids, crosscap coefficients are directly related to the compactification radius and can be used to extract the Luttinger parameter~\cite{TanBY2025}.

\vspace{1em}

{\it Data availability} --- The data that support the findings of this article are openly available at~\cite{DataLink}.

\vspace{1em}

{\it Acknowledgments} --- We are very grateful to Lei Wang for previous collaborations on related problems. We thank Wei Zhu for helpful discussions and the authors of DiagHam for sharing their programs~\cite{DiagHam}. This work was supported by NNSF of China under Grant No. 12174130 (Y.H.W) and the Sino-German (CSC-DAAD) Postdoc Scholarship Program (Y.S.Z.).

\bibliography{ReferConde}

\end{document}